%% file: article.tex
\documentclass[11pt,a4paper]{article}
\usepackage[margin=3cm]{geometry}

\usepackage{subcaption}
\usepackage{mathtools}
\usepackage{amssymb}
\usepackage{amsthm}
\usepackage{cancel}
\usepackage{xfrac}
\usepackage{multirow}

\newcommand{\transpose}{\mathsf{T}}
\newcommand{\zeromat}[2]{\boldsymbol{0}_{#1 \times #2}}
\newcommand{\onemat}[2]{\boldsymbol{1}_{#1 \times #2}}
\newcommand{\figref}[1]{Figure~\ref{#1}}

\newtheorem{rank-equality}{Theorem}
\newtheorem{link-equivalence}{Lemma}

\makeindex

\title{Black Hole Metric: Overcoming the PageRank Normalization Problem}

\author{M.~Buzzanca \and V.~Carchiolo \and A.~Longheu \and M.~Malgeri \and G.~Mangioni\thanks{Corresponding Author: giuseppe.mangioni@dieei.unict.it} \\
		\mbox{}\\
    DIEEI, University of Catania, ITALY
}

\date{}

\begin{document}
\maketitle

\begin{abstract}
    In network science, there is often the need to sort the graph nodes. While
    the sorting strategy may be different, in general sorting is performed
    by exploiting the network structure. In particular, the metric PageRank has
    been used in the past decade in different ways to produce a ranking based
    on how many neighbors point to a specific node. PageRank is simple, easy
    to compute and effective in many applications, however it comes with a
    price: as PageRank is an application of the random walker, the arc weights
    need to be normalized. This normalization, while necessary,
    introduces a series of unwanted side-effects. In this paper, we propose a
    generalization of PageRank named Black Hole Metric which mitigates the
    problem. We devise a scenario in which the side-effects are particularily
    impactful on the ranking, test the new metric in both real and synthetic
    networks, and show the results.
\end{abstract}

\input{introduction}
\input{related-works}
\input{pagerank}

\input{black-hole}
\input{experiments}
\input{conclusion}

\section*{Acknowledgements}
Funding: This work was supported by S.M.I.T. Sistema di Monitoraggio
integrato per il Turismo -- Linea di intervento 4.1.1.1 del POR FESR Sicilia
2007-2013.


\end{document}

%% file: introduction.tex
\section[Introduction]{Introduction} \label{s:introduction}
In the vast amount of digital data, humans have the need to discriminate those
relevant for their purposes to effectively transform them into useful
information, which usefulness depends on the scenario being considered. For
instance, in web searching we aim at finding significant pages with respect to
an issued query~\cite{Roa-Valverde:2014}, in an E-learning context we look for
useful resources within a given
topic~\cite{Carchiolo20101893,serrano:2011,serrano:2013}, or in a
recommendation network we search for most reliable entities to interact
with~\cite{Chen:2015,Carchiolo2015,kim:2012,bedi:2014}. All these situations
fall under the umbrella of \emph{ranking}, a challenge addressed in these years
through different solutions. The most well-known technique is probably the
PageRank algorithm~\cite{pagerank:98,pagerank:99}, originally designed to be
the core of the Google ({\em www.google.com}) web search engine. Since it was
published it has been
analyzed~\cite{pretto:2002,Bianchini:2005,Langville:2004}, modified or extended
for use in other contexts~\cite{zhirov:2010,Gupta:2013}, to overcome some of
its limitations, and to address computational
issues~\cite{lee:2003,richardson:2002}.

PageRank has been widely adopted in several different application scenarios. In
this paper, we propose a generalization of PageRank whose motivation stems from
the concept of trust in virtual social networks. In this context trust is
generally intended as a measure of the assured reliance on a specific feature
of someone~\cite{marsh:1994,Mcknight:1996,artz@2007}, and it is exploited to
rank participants in order to discover the \emph{best} entities that is "safe"
to interact with. This trust-based ranking approach allows to cope with
uncertainty and risks~\cite{ruohomaa:2005}, a feature especially relevant in
the case of lack of bodily presence of counterparts.

\begin{figure}[htpb]
    \centering
    \begin{subfigure}{0.4\textwidth}
        \includegraphics[width=\textwidth]{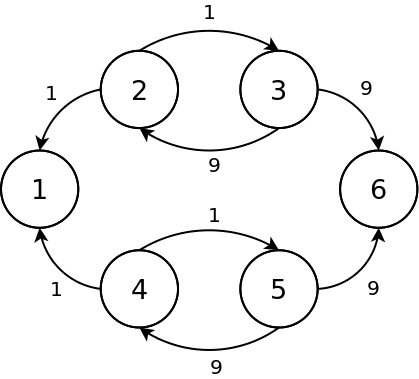}
        \caption{Before the normalization}
        \label{fig:sample-1}
    \end{subfigure}
    \qquad
    \begin{subfigure}{0.4\textwidth}
        \includegraphics[width=\textwidth]{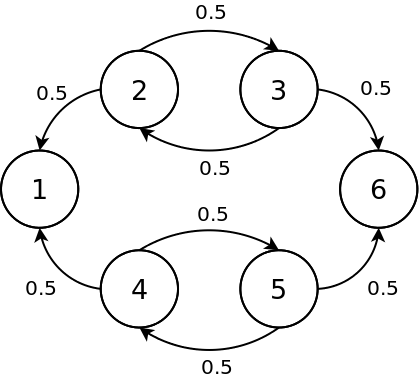}
        \caption{After the normalization}
        \label{fig:sample-1-pr}
    \end{subfigure}
    \caption{Network with asymmetric trust distribution}
    \label{fig:sample-1-both}
\end{figure}

A notable limitation of PageRank when it's used to model social behaviour, is
its inability to preserve the absolute arc weights due to the normalization
introduced by the application of the random walker. In order to illustrate the
problem, we introduce a weighted network where arcs model relationships among
entities. Entities may be persons, online shops, computers that
in general need to establish relationships with other entities of the same
type. Let's suppose to have the network shown in \figref{fig:sample-1}, where each
arc weight ranges over $[0, 10]$.

Given the network topology, intuition suggests that node 1 would be regarded
more poorly compared to node 6 since it receives lower trust values from his neighbors, but, as detailed later, normalizing the weights
alters the network topology so much that both nodes are placed in the same
position in the ranking. The normalization of the outlink weights indeed
hides the weight distribution asymmetry, as depicted in
\figref{fig:sample-1-pr}.

Moreover, PageRank shadows the social implications of assigning low weights to
all of a node's neighbours. If we consider the arcs as if they were social
links, common sense would tell us to avoid links with low weight, as they
usually model worse relationships. If we look at the normalized weights in
\figref{fig:sample-1-pr} though, we can see that in many cases, the normalized
weight changes the relationship in a counter-intuitive way. Consider the arcs
going from node 2 or 4 to their neighbours: we can see that their normalized
weights are set to 0.5, which, in the range $[0, 1]$ is an average score.
However, the original weight of those links was 1, a comparatively lower score
considering that the original range was $[0, 10]$.

The contribution of this paper is the proposal of a new PageRank-based metric
we name \emph{Black Hole Metric} to cope with the normalization effect and to
deal with the issue of the skewed arc weights, detalied in Section
\ref{s:black-hole}. Note that our proposal seamlessly adapts
to any situation where PageRank can be used, being not limited to trust networks; in the following, they are considered as a simple case study.

The paper is organized as follows: in section ~\ref{s:related-works} we give an
overview of the existing literature concerning PageRank, in
section~\ref{s:pagerank} we describe the PageRank algorithm together with some
of its extensions, whereas in section~\ref{s:black-hole} we illustrate our
proposal in detail, comparing it to the basic PageRank in
section~\ref{s:experiments}, where we also show some first experiments.
Finally, in section~\ref{s:conclusion} we provide our conclusions and some open
discussions.

%% file: related-works.tex
\section[Related Works]{Related Works} \label{s:related-works}
PageRank is essentially an application of the random walker model on a Markov
chain: the nodes of the Markov chain are the web pages, and the arcs are the
links that connect one page to another. The walker represents a generic web
surfer which moves from page to page with a certain probability, according to
the network structure, and occasionally "gets bored" and jumps to a random node
in the network. The steady-state probability vector of the random walker
process holds the PageRank values for each node, which can be used to determine
the global ranking.

Before describing in detail the normalization problem by showing the issues
that may occur, we briefly introduce the
web ranking metric that would lay the foundation to modern search engines. As
previously mentioned, the PageRank algorithm has been thoroughly analysed, both
in its merits and in its shortcomings. Although Pagerank was proposed a long
time ago, it still lives as the backbone of several technologies, not limited
to the web domain. For example, in \cite{Gupta:2013}, personalized PageRank is
cited as a possible algorithm to be used in Twitter's "Who To Follow"
architecture. In \cite{Gleich:2014}, the author shows how the mathematics
behind PageRank have been used in a plethora of applications which are not
limited to ranking pages on the web. In \cite{Senanayake:2015}, another
PageRank extension appears as a tentative replacement of the h-index for
publications.  A recent technology that uses personalized PageRank as its
backend is SwiftType (https://swiftype.com/) which is gaining popularity as a
search engine for various platforms.

These examples, however, do not use the basic version of PageRank, but some
sort of extension that better fits the domain in which it is applied.  To the
best of our knowledge, the original PageRank algorithm is seldom used "as it
is", our proposal itself is an alternative to the weighted PageRank algorithm
described in \cite{Wenpu:2004}. Besides our work, there are several other
extensions that were proposed in the past. For example, CheiRank
\cite{zhirov:2010} focuses on evaluating the outlink strength instead of the
inlink strength, with the ultimate effect of rewarding hub behaviour, making it
essentially a dual metric of PageRank. DirichletRank \cite{Wang:2008} is a
derivative metric which claims to solve the "zero-one gap problem" of the
original PageRank (in brief, the teleportation chance drops from $1$ to $d$,
where $d < 1$, when the number of outlinks change from 0 to 1).
\cite{richardson:2002} proposes a query-based PageRank in which the random
walker probabilities are dependent on the query relevance.

These variants are essentially alternatives or improvements, but several other
works either focus on providing shorter runtime, or are adaptations of PageRank
to different domains. For instance, \cite{Bahmani:2010} proposes a Monte Carlo
technique to perform fast computation of random walk based algorithms such as
PageRank. There also exists at least a version of distributed PageRank
\cite{Zhu:2005} which better handles the ever increasing number of web pages to
rank: one of the main shortcoming of basic PageRank is the inability of holding
large link matrices entirely in main memory, resulting in slowed down I/O
operation; the distributed version of the algorithm takes care of this problem.
Another approach \cite{Bahmani:2011} handles large data samples by making use
of the MapReduce algorithm, introducing PageRank to the world of Big Data.
Other types of optimization methods can be found in the surveys
\cite{Berkhin:2005} and \cite{Sargolzaei:2010}. An important extension of
PageRank that operates on the trust network domain\cite{CPE:CPE1856}\cite{Carchiolo2012}\cite{Carchiolo:2013}, 
is EigenTrust \cite{kamvar:2003}. Since the application scenario involves trust networks, the
entities involved change slightly: web pages are replaced with network nodes,
links are replaced with arcs. The underlying mathematics, however, remain
essentially unchanged.

PageRank and its extensions have to face a plethora of competitors in several
application domains. In the web domain we have HITS \cite{kleinberg:1998},
which is not based on the random walker model and is able to provide both an
"authority" ranking, which rewards nodes that have many backlinks, and a "hub"
ranking, which rewards nodes that have many forward links. SALSA
\cite{Lempel:2001} computes a random walk on the network graph, but integrates
the search query into the algorithm, which is something PageRank does not do.
In the trust networks domain we have PeerTrust \cite{Xiong:2004}, which
computes the global trust by aggregating several factors, and PowerTrust
\cite{zhou:tpds:2007} which uses the concept of "power nodes", which are
dynamically selected, high reliability nodes, that serve as moderators for the
global reputation update process. The PowerTrust article also describes how the
algorithm compares to EigenTrust with a set of simulations that analyse its
performance. Several articles feature side-by-side comparisons among PageRank
(and its extensions) and other metrics
\cite{Ashutosh:2009,Prabha:2014,Najork:2007}. In particular,
\cite{HyunChul:2003} focuses on comparing HITS, PageRank and SALSA, and its
authors prove that PageRank is the only metric that guarantees algorithmic
stability with every graph topology.

%% file: pagerank.tex
\section[PageRank]{PageRank} \label{s:pagerank}

\subsection{Definitions and Notation} \label{ss:def-notation}
In order to better understand the mathematics of the Black Hole Metric, we need
to clarify the notation used throughout this paper and provide a few
definitions, which are similar to the notations used in the article of
PageRank. Let us suppose that $N$ is the number of nodes in the network. We
will call $A$ the $N \times N$ \emph{network adjacency matrix} or \emph{link
    matrix}, where each $a_{ij}$ is the weight of the arc going from node $i$
to node $j$. $S$ is the $N \times 1$ \emph{sink vector}, defined as:
\[
	s_{i} = \begin{cases}
		1 & \mbox{if } {out}_{i} = 0 \\
		0 & \mbox{otherwise}
	\end{cases}
    \quad \forall i \le N
\]

where ${out}_i$ is the number of outlinks of node $i$. $V$ is the
\emph{personalization vector} of size $1 \times N$, equal to the transposed
initial distribution probability vector in the Markov chain model
$P_0^\transpose$. While this vector can be arbitrarily chosen as long as it's
stochastic, a common choice is to make each term equal to $\sfrac{1}{N}$. $T =
\onemat{N}{1}$ is the \emph{teleportation vector}, where the notation
$\onemat{N}{M}$ stands for a $N \times M$ matrix where each element is $1$.

In the general case the Markov chain built upon the network graph is not always
ergodic, so it is not used directly for the calculation of the steady state
random walker probabilities. As described in \cite{pagerank:99}, the
\emph{transition matrix} $M$, used in the associated random walker problem, is
derived from the link matrix, the sinks vector, the teleportation vector and
the personalization vector defined above:
\begin{equation} \label{eq:m-normal}
    M = d (A + SV) + \left(1 - d\right) TV
\end{equation}
where $d \in [0, 1]$ is called \emph{damping factor} and it is commonly set to
$0.85$. As we know from the Markov chain theory, the random walk probability
vector at step $n$ can be calculated as:
\begin{equation} \label{eq:pn-normal}
	P_n = M^\transpose P_{n-1}
\end{equation}
the related random walker problem can be calculated as:
\begin{equation} \label{eq:rw-normal}
	P = \left( \lim_{n \to \infty} M^n \right)^\transpose P_0 = \lim_{n \to
        \infty}(M^\transpose)^n P_0 =M_\infty^\transpose P_0
\end{equation}

\subsection{The normalization problem}
Let us calculate the PageRank values of the sample network in
\figref{fig:sample-1} to highlight the flattening effect of the normalization.
By applying the definitions in section \ref{ss:def-notation} the network in
\figref{fig:sample-1-pr} can be described by the following matrices and
vectors:
\[
	\setlength\arraycolsep{2pt}
	A =
	\begin{pmatrix}
		0	& 0		& 0		& 0		& 0		& 0		\\[0.3em]
		0.5	& 0		& 0.5	& 0		& 0		& 0		\\[0.3em]
		0	& 0.5	& 0		& 0		& 0		& 0.5	\\[0.3em]
		0.5	& 0		& 0		& 0		& 0.5	& 0		\\[0.3em]
        0	& 0		& 0		& 0.5	& 0		& 0.5	\\[0.3em]
		0	& 0		& 0		& 0		& 0		& 0
	\end{pmatrix}
	\quad
	S =
	\begin{pmatrix}
		1 \\[0.3em]
		0 \\[0.3em]
		0 \\[0.3em]
		0 \\[0.3em]
		0 \\[0.3em]
		1
	\end{pmatrix}
	\quad
	V = P_0^\transpose = \frac{1}{6} \cdot \onemat{1}{6}
	\quad
	T = \onemat{6}{1}
\]
If we calculate the PageRank values for the nodes of the sample network
assuming $d = 0.85$ we obtain:
\[
	p_1 = p_6 = 0.208 \quad p_2 = p_3 = p_4 = p_5 = 0.146
\]
Note that the nodes 1 and 6 are both first in global ranking, despite the fact
that their in-strength was so different before the normalization.

%% file: black-hole.tex
\section[Black Hole Metric]{Black Hole Metric} \label{s:black-hole}
In order to avoid the flattening effect of the PageRank normalization, we
propose a new metric named Black Hole Metric. Black Hole Metric globally
preserves the proportions among the arc weights, and ensures at the same time
that the outstrength is equal to 1 for each node. This allows compatibility
with the random walker model, and it is done by applying a transformation to
the original network. The transformation only requires the knowledge of the
maximum and the minimum value each weight can assume. This range bounds may
be global (each node has the same scale) or local (each node has its own weight
scale); in practice, global scale is preferred.

\begin{figure}[htpb]
    \centering
    \begin{subfigure}{0.25\textwidth}
        \includegraphics[width=\textwidth]{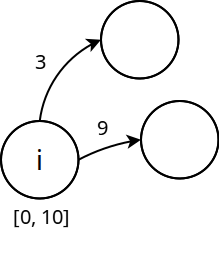}
    \end{subfigure}
    \qquad
    \begin{subfigure}{0.25\textwidth}
        \includegraphics[width=\textwidth]{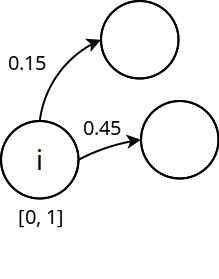}
    \end{subfigure}
    \qquad
    \begin{subfigure}{0.25\textwidth}
        \includegraphics[width=\textwidth]{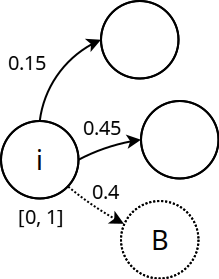}
    \end{subfigure}
    \caption{Transformation steps}
    \label{fig:transformation}
\end{figure}

At this stage, we will provide an example of the transformation steps as illustrated
in \figref{fig:transformation}. In order to obtain the depicted values, we used
formulas \eqref{eq:bh-arc-weight} and \eqref{eq:bh-weight}, which will be
explained in detail in paragraph \ref{ss:weight-assignment}. Before tackling
the mathematical part though, we will now describe qualitatively how the Black
Hole Metric operates. First, it changes the original weights so that they lie
in the range $[0, 1]$. The resulting outstrength $s_i$ of each node $i$ is not
preserved, but it is guaranteed to be less or equal than 1.  Then, we
introduce a new node, the \emph{black hole}, and node $i$ is connected to it. The
strength of this connection is set to $1 - s_i$, as if the black hole
"absorbed" the missing weight amount to reach 1 as the total $i$'s outstrenght. 
This transformation is applied to all nodes in the network.

Since the black hole does not have outlinks, it is a sink by construction, and
the random walker can only move away because of the teleportation effect. In a
network without the black hole, each node would normally have a $1 - d$ chance
to teleport to a random node instead of going towards one of its neighbours. We
know that moving to the black hole from node $i$ occurs with a $1 - s_i$
chance, and that once in the black hole, the random walker inevitably teleports
to a random node. In conclusion, taking both effects into account, each node
has a $(1 - d)(1 - s_i)$ chance to teleport to a random node, where $d$ is the
damping factor as in \eqref{eq:m-normal}. 

It is important to note that not every network has a defined scale for its arc
weights. There are networks in which the weights are \emph{unbounded}: an
example would be an airline transportation network in which each arc weight is
the number of flights connecting two cities. As there is no real "maximum",
there is no trivial weight scale that can be used for the transformation. In
order to apply such transformation in an unbounded network, we need to somehow
infer meaningful scale boundaries exploiting the knowledge of the domain, and
the topology of the network. This is usually non-trivial and for the sake of
simplicity, in this paper we take into account only bounded networks.

\subsection{Weight assignment} \label{ss:weight-assignment}
In this section, we explain how the new weights are calculated. Let $i$ be a
generic node in the network. Let the interval $[l_i, h_i]$ be the \emph{local}
scale of node $i$. Let $r_{ij}$ be the weight that node $i$ assigns to the arc
pointing towards node $j$. Let ${out}_i$ be the number of neighbours of node
$i$. Given that $l_i \le r_{ij} \le h_i$, We define the modified weight
$\bar{a}_{ij}$ of the arc that goes from $i$ to $j$ as:
\begin{equation} \label{eq:bh-arc-weight}
    \bar{a}_{ij} = \frac{r_{ij} - l_i}{{out}_i (h_i - l_i)}
\end{equation}
which is significantly different from the normalized arc weight required by
PageRank:
\begin{equation} \label{eq:pr-arc-weight}
    a_{ij} =  \frac{r_{ij}}{\sum_{k=1}^{{out}_i} r_{ik}}
\end{equation}
As mentioned before, the resulting node outstrength is only guaranteed to be
less or equal to 1:
\begin{equation} \label{eq:aij-nostochastic}
    \sum_{j=1}^{{out}_i} \bar{a}_{ij} = \sum_{j=1}^{{out}_i} \frac{r_{ij} -
        l_i}{{out}_i (h_i - l_i)} \le \sum_{j=1}^{{out}_i}
    \frac{1}{{out}_i} = 1
\end{equation}
We purposely excluded the contribute of the arc from node $i$ to the black hole
in \eqref{eq:aij-nostochastic}, which is:
\begin{equation} \label{eq:bh-weight}
    b_i = \sum_{j=1}^{{out}_i } \frac{h_i - r_{ij}}{{out}_i (h_i - l_i)}
\end{equation}
If we include this contribute as well, the weight sum becomes 1 as desired:
\begin{equation} \label{eq:bh-stochastic}
    \sum_{j=1}^{{out}_i} \bar{a}_{ij} + b_i =  \sum_{j=1}^{{out}_i}
    \frac{r_{ij} - l_i}{{out}_i (h_i - l_i)} + \sum_{j=1}^{{out}_i} \frac{h_i -
        r_{ij}}{{out}_i (h_i - l_i)} = \sum_{j=1}^{{out}_i} \frac{h_i -
        l_i}{{out}_i (h_i - l_i)} = 1
\end{equation}
The weight $b_i$ is ultimately the probability that the node would rather visit
a random node rather than one of its neighbours, which is the \emph{amplification} of
the teleportation effect operated by the network transformation described
before.

\subsection{Proposal} \label{ss:formal-proposition}
With the previously mentioned weight assignment, it is now possible to define
Black Hole Metric as a generalization of PageRank. Let's start by defining the
new link matrix $A'$, the new sink vector $S'$, the new teleportation vector
$T'$ and the new personalization vector $V'$.

For the sake of convenience, we name $B$ the \textit{black hole vector}, which
is the $N \times 1$ vector that holds the weights of the arcs going from each
node to the black hole. The updated link matrix $A'$ is obtained by combining
the $B$ vector with $\bar{A} = \{ \bar{a}_{ij} \}$ where $\bar{a}_{ij}$ is
defined in \eqref{eq:bh-arc-weight}:
\begin{equation} \label{eq:a-bh}
	A' =
	\begin{pmatrix}
		\bar{A}	        & B \\[0.3em]
		\zeromat{1}{N}  & 0
	\end{pmatrix}
\end{equation}
In general, $A \ne \bar{A}$. There are other three entities involved in the
computation of the transition matrix used by the random walker model: the
teleportation vector $T'$, the personalization vector $V'$, and the sink vector
$S'$. We may define $T'$ and $V'$ as follows:
\begin{equation} \label{eq:vt-bh}
	V' = {P'_0}{}^\transpose =
	\begin{pmatrix}
		V & 0
	\end{pmatrix} =
	\begin{pmatrix}
		\frac{1}{N} \cdot \onemat{1}{N} & 0
	\end{pmatrix}
	\quad
	T' =
	\begin{pmatrix}
		T \\[0.3em]
        0
	\end{pmatrix} =
	\begin{pmatrix}
		\onemat{N}{1} \\[0.3em]
		0
	\end{pmatrix}
\end{equation}
Note that we deliberately excluded the \emph{black hole} from the teleportation
effect by putting a value of $0$ in the corresponding entries of $T'$ and $V'$.
Since the black hole is a sink by construction, going back there as the
consequence of a teleportation effect would only trigger another teleportation
effect, which is unnecessary.

Regarding the sink vector, we intuitively want to set to $1$ the
corresponding index in the vector, as the black hole is a sink, but this
actually makes the black hole row in the link matrix not stochastic. Let's
consider the matrix $A'$ defined above, and let's use the following sink vector
to compute the transition matrix:
\begin{equation} \label{eq:s-star}
    S^* =
    \begin{pmatrix}
        S \\[0.3em]
        1
    \end{pmatrix}
\end{equation}
We can calculate $M'$ using \eqref{eq:m-normal}:
\[
    \begin{split}
        M' = d (A' + S^*) + (1 - d)T'V' = d \left[
            \begin{pmatrix}
                \bar{A} & B \\[0.3em]
                \zeromat{1}{N} & 0
            \end{pmatrix} +
            \begin{pmatrix}
                SV & \zeromat{N}{1} \\[0.3em]
                V & 0
            \end{pmatrix}
        \right] + \\
        + (1 - d)
        \begin{pmatrix}
            TV & \zeromat{N}{1} \\[0.3em]
            \zeromat{1}{N} & 0
        \end{pmatrix} =
        \begin{pmatrix}
            d(\bar{A} + SV) + (1 - d) TV & dB \\[0.3em]
            dV & 0
        \end{pmatrix}
    \end{split}
\]
The black hole row in the link matrix is $dV$, which is not stochastic: the
vector $V$ is, but since $d \ne 1$ the product is not. This happened because we
excluded the black hole from the teleportation effect by setting its entry to
$0$ in $T'$, which interferes with the damping factor correction. In order to
compensate for this effect, it is sufficient to multiply the black hole entry
in the sink vector by a $\frac{1}{d}$ term:
\begin{equation} \label{eq:s-bh}
    S' =
    \begin{pmatrix}
        S           \\[0.3em]
        \frac{1}{d}
    \end{pmatrix}
\end{equation}
this makes the black hole row in the link matrix $V$, which is stochastic.
Equations \eqref{eq:a-bh}, \eqref{eq:vt-bh} and \eqref{eq:s-bh} allow us to
define the random walker model according to the definition of $M$ in
\eqref{eq:m-normal}:
\[
    M' = d (A'+ S'V') + \left(1 - d\right) T'V'
\]
We can now partition $M'$:
\[
\begin{split}
M' = d
	\begin{pmatrix}
		\bar{A}		    & B \\[0.3em]
		\zeromat{1}{N}  & 0
	\end{pmatrix} + d
	\begin{pmatrix}
		S \\[0.3em]
		\frac{1}{d}
	\end{pmatrix}
	\begin{pmatrix}
		V & 0
	\end{pmatrix} + (1 - d)
	\begin{pmatrix}
		T \\[0.3em]
		0
	\end{pmatrix}
	\begin{pmatrix}
		V & 0
	\end{pmatrix} =
	\begin{pmatrix}
		d \bar{A}	    & dB \\[0.3em]
		\zeromat{1}{N}	& 0
	\end{pmatrix} + \\
    + \begin{pmatrix}
		dSV	& \zeromat{N}{1} \\[0.3em]
		V	& 0
	\end{pmatrix} +
	\begin{pmatrix}
		(1 - d) TV		& \zeromat{N}{1} \\[0.3em]
		\zeromat{1}{N}	& 0
	\end{pmatrix} =
	\begin{pmatrix}
		d \left(\bar{A} + SV\right) + (1 - d) TV	& dB \\[0.3em]
		V											& 0
	\end{pmatrix}
\end{split}
\]
If we name $\bar{M} = d \left(\bar{A} + SV\right) + (1 - d) TV$ we have:
\begin{equation} \label{eq:m-bh}
M' = \begin{pmatrix}
    \bar{M} & dB \\[0.3em]
		V & 0
	\end{pmatrix}
\end{equation}
Consider now the following partition of the rank vector $P'$:
\begin{equation} \label{eq:p-bh}
	P' =
	\begin{pmatrix}
		\bar{P}		\\[0.3em]
		p_b
	\end{pmatrix}
\end{equation}
where $p_b$ is the steady-state probability of the black hole. Note that
usually $\bar{P} \neq P$. The rank vector at step $n$, which we named $P'_n$,
can be obtained using \eqref{eq:pn-normal}, \eqref{eq:m-bh} and
\eqref{eq:p-bh}:
\[
\begin{split}
P'_n = M'{}^\transpose P'_{n-1} \Leftrightarrow
	\begin{pmatrix}
		\bar{P}_n \\[0.3em]
		p_{b_n}
	\end{pmatrix} =
	\begin{pmatrix}
        \bar{M}^\transpose 	& V^\transpose \\[0.3em]
		dB^\transpose						& 0
	\end{pmatrix}
	\begin{pmatrix}
		\bar{P}_{n-1} \\[0.3em]
		p_{b_{n-1}}
	\end{pmatrix} \Leftrightarrow
	\begin{pmatrix}
		\bar{P}_n \\[0.3em]
		p_{b_n}
	\end{pmatrix} = \begin{pmatrix}
        \bar{M}^\transpose \bar{P}_{n-1} + p_{b_{n-1}}
            V^\transpose \\[0.3em]
        dB^\transpose \bar{P}_{n-1}
\end{pmatrix}
\end{split}
\]
We split the calculation in two parts:
\begin{equation} \label{eq:pn-bh}
	\begin{cases}
		\bar{P}_n = \bar{M}^\transpose \bar{P}_{n-1} +
            p_{b_{n-1}} V^\transpose \\
		p_{b_n} = d B^\transpose \bar{P}_{n-1}
	\end{cases}
\end{equation}
The related random walker process \eqref{eq:rw-normal}, given the definition of
matrix $M'$ \eqref{eq:m-bh}, the definition of the personalization vector $P'_0
=V'{}^\transpose$, and the definition of the rank vector of the Black Hole
Metric $P'$ \eqref{eq:p-bh}, can be written as:
\begin{equation} \label{eq:rw-bh}
	P' = M'_\infty{}^\transpose P'_0 \Leftrightarrow
	\begin{pmatrix}
		\bar{P}	\\[0.3em]
		p_b
	\end{pmatrix} =
	\begin{pmatrix}
        \bar{M}^\transpose	& V^\transpose \\[0.3em]
		d B^\transpose						& 0
	\end{pmatrix}_\infty
	\begin{pmatrix}
		P_0	\\[0.3em]
		0
	\end{pmatrix}
\end{equation}

An important property of the transition matrix $M'$ is that it leads to a
converging random walker process no matter the network topology, as it will be
clarified in section \ref{ss:proof-conv}. As a final note, even though in
general $A \neq \bar{A}$ and $P \neq \bar{P}$, in section \ref{ss:rank-eq} we
will introduce a sufficient condition that allows the identity.

\subsection{Application to example toy network} \label{ss:example}
\begin{figure}[htpb]
	\centering
	\includegraphics[width=0.55\textwidth]{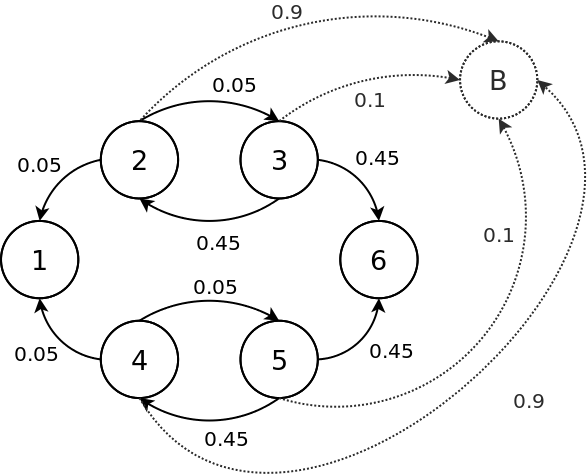}
    \caption{The Network in \figref{fig:sample-1-both} with the black
        hole.}
	\label{fig:sample-1-bh}
\end{figure}

Now that we have defined the necessary entities and described how we assign
weights in the modified network, let's see how Black Hole Metric behaves in the
sample trust network in \figref{fig:sample-1}. For this particular network we
set that $l_i = l = 0, \enspace h_i = h = 10 \enspace \forall i \in [1,N]$. It
is easy to note that we have only three types of nodes in the network:
\begin{enumerate}
	\item Nodes which have two links with weight 1 out of 10 (nodes 2 and 4).
	\item Nodes which have two links with weight 9 out of 10 (nodes 3 and 5).
	\item Sinks (nodes 1 and 6).
\end{enumerate}
We only show the arc weights of node 2, as the same formulas can be used to
calculate the outlink weights of the other nodes. Given that ${out}_2 = 2$ we
have:
\[
    \bar{a}_{21} = \bar{a}_{23} = \frac{r_{21} - l}{{out}_2 \cdot (h -
        l)} = \frac{1 - 0}{2 \cdot (10 - 0)} = \frac{1}{20}
\]
The black hole arc weight is going to be:
\[
    b_2 = \frac{h - r_{21} + h - r_{23}}{{out}_2 (h - l)} = \frac{20 -
        2}{2 \cdot (10 - 0)} = \frac{9}{10}
\]
as expected, $\bar{a}_{21} + \bar{a}_{22} + b_2 = 1$. By applying the
formulas to all arcs we create the network in \figref{fig:sample-1-bh}.
The link matrix $A'$ as in \eqref{eq:a-bh} is:
\[
	\setlength\arraycolsep{2pt}
	A' =
	\begin{pmatrix}
		0		& 0		& 0		& 0	    & 0		& 0		& 0     \\[0.3em]
		0.05	& 0		& 0.05	& 0	    & 0		& 0		& 0.9   \\[0.3em]
		0		& 0.45	& 0		& 0	    & 0		& 0.45	& 0.1	\\[0.3em]
		0.05	& 0		& 0		& 0	    & 0.05	& 0		& 0.9	\\[0.3em]
		0		& 0		& 0		& 0.45  & 0		& 0.45	& 0.1	\\[0.3em]
		0		& 0		& 0		& 0	    & 0		& 0		& 0 	\\[0.3em]
		0		& 0		& 0		& 0	    & 0		& 0		& 0
	\end{pmatrix}
\]
while $B$ is:
\[
B =
\begin{pmatrix}
	0	\\[0.3em]
	0.9	\\[0.3em]
	0.1	\\[0.3em]
	0.9	\\[0.3em]
	0.1	\\[0.3em]
	0	\\[0.3em]
\end{pmatrix}
\]
Vectors $S'$, $V'$ and $T'$ are obvious from \eqref{eq:vt-bh} and
\eqref{eq:s-bh}:
\[
	S' =
	\begin{pmatrix}
		1 \\[0.3em]
		0 \\[0.3em]
		0 \\[0.3em]
		0 \\[0.3em]
		0 \\[0.3em]
		1 \\[0.3em]
		\frac{1}{d}
	\end{pmatrix}
    \quad
	V' = {P'_0}{}^\transpose =
	\begin{pmatrix}
		\frac{1}{6} \cdot \onemat{1}{6} & 0
	\end{pmatrix}
	\quad
	T' =
	\begin{pmatrix}
		\onemat{6}{1} \\[0.3em]
		0
	\end{pmatrix}
\]
If we compute the steady-state probabilities for the random walker process in
\eqref{eq:rw-bh} assuming $d = 0.85$, the values calculated for each node
(including the black hole) of the network in \figref{fig:sample-1-bh} are:
\[
p_1 = 0.110 \quad p_2 = p_4 = 0.138 \quad p_3 = p_5 = 0.104 \quad p_6 = 0.178
    \quad p_b = 0.228
\]
which better models the trust relationships among the nodes, as $p_1 < p_6$.
There is also a value $p_b$ for the black hole, which is a consequence of the
transformation we operated. Since the black hole is not a real node, this
probability does not bear any particular meaning, and it can be discarded.

\subsection{Complexity assessment}
Using \eqref{eq:pn-bh} for direct computation, no matter the method in use, is
inefficient in both time and space complexity, therefore we will now introduce
a more efficient way to solve the problem. First, let us rewrite $\bar{P}_n$
appropriately:
\[
    \bar{P}_n = \bar{M}^\transpose \bar{P}_{n-1} + p_{b_{n-1}} V^\transpose =
    d\bar{A}^\transpose \bar{P}_{n-1} + d V^\transpose S^\transpose
    \bar{P}_{n-1} + \left(1 - d\right) V^\transpose T^\transpose \bar{P}_{n-1}
    + p_{b_{n-1}} V^\transpose
\]
The quantities $T^\transpose \bar{P}_{n-1} = \bar{t}_{p_{n-1}}$ and $
S^\transpose \bar{P}_{n-1} = \bar{s}_{p_{n-1}}$ are both scalars. In
particular, we have:
\begin{equation} \label{eq:tpn-bh}
 T^\transpose \bar{P}_{n-1} = \sum_{k=0}^N{\bar{p}_{k_{n-1}}} = 1 - p_{b_{n-1}}
\end{equation}
which allows us to write:
\[
\begin{split}
    \bar{P}_n = d\bar{A}^\transpose \bar{P}_{n-1} + d \bar{s}_{p_{n-1}}
    V^\transpose + \left(1 - d\right) \bar{t}_{p_{n-1}} V^\transpose
    + p_{b_{n-1}} V^\transpose = \\
    = d\bar{A}^\transpose \bar{P}_{n-1} + [d \bar{s}_{p_{n-1}}
    + \left(1 - d\right) \bar{t}_{p_{n-1}} + p_{b_{n-1}}] V^\transpose
\end{split}
\]
The quantity under square brackets can be further simplified using
\eqref{eq:tpn-bh}:
\[
	d \bar{s}_{p_{n-1}} + \left(1 - d\right) \bar{t}_{p_{n-1}} + p_{b_{n-1}} =
	d \bar{s}_{p_{n-1}} + (1 - d) (1 - p_{b_{n-1}}) + p_{b_{n-1}} =
\]
\[
	= d \bar{s}_{p_{n-1}} + 1 - d - \cancel{p_{b_{n-1}}} + d p_{b_{n-1}} +
    \cancel{p_{b_{n-1}}} = 1 - d (1 - \bar{s}_{p_{n-1}} - p_{b_{n-1}})
\]
which allows us to write \eqref{eq:pn-bh} as:
\begin{equation} \label{eq:pn-simple-bh}
	\begin{cases}
		\bar{P}_n = d \bar{A}^\transpose \bar{P}_{n-1} +
            [1 - d (1 - \bar{s}_{p_{n-1}} - p_{b_{n-1}})] V^\transpose \\
		p_{b_n} = d \bar{b}_{p_{n-1}}
	\end{cases}
\end{equation}
$\bar{b}_{p_{n-1}} = B^\transpose \bar{P}_{n-1}$ is also a scalar. The index
form of \eqref{eq:pn-simple-bh} is:
\[
	\begin{cases}
		\bar{p}_{i_n} = d \sum_{h = 0}^N\limits \bar{a}_{hi}
            \bar{p}_{h_{n-1}} + [1 - d (1 - \bar{s}_{p_{n-1}} -
            p_{b_{n-1}}) ] v_i \\
		p_{b_n} = d \bar{b}_{p_{n-1}}
	\end{cases}
\]

There are three expensive computations in \eqref{eq:pn-simple-bh}, which
complexity is easily inferrable:
\begin{itemize}
    \item $\bar{A}^\transpose \bar{P}_{n-1}$. Matrix by vector products usually
        have a computational complexity of $O(N^2)$. However, in our case, we
        know that matrix $\bar{A}$ has very few non-zero entries. This number
        is equal to $|E|$, the total number of arcs in the network, so we can
        conclude that the average computational complexity is $O(|E|)$ which is
        less than $O(N^2)$ in the general case.
    \item $\bar{s}_{p_{n-1}} = S^\transpose \bar{P}_{n-1}$ and
        $\bar{b}_{p_{n-1}} = B^\transpose \bar{P}_{n-1}$. Inner products among
        vectors always have complexity $O(N)$. $N$ is less than $|E|$,
        unless the overall number of arcs in the network is less than the
        number of nodes itself, which seldom happens.
\end{itemize}
Then, the overall complexity is $O(|E|)$ in the average case,
which is the same as PageRank. Note that $T^\transpose\bar{P}_{n-1}$ does not
add to the complexity, as it can be written as the scalar $1 - p_{b_{n-1}}$ and
computed offline.

Furthermore, we analyse the memory usage of the entities involved outside the
computation:
\begin{itemize}
    \item Memory usage for adjacency sparse matrix $\bar{A}$ depends
        on how it is stored. Assuming the storage format is Compressed Column
        Storage, it is proportional to $2|E| + N + 1$.
    \item Memory usage for personalization vector $V$, sink vector $S$ and
        black hole vector $B$ is proportional to $N$.
    \item No memory usage for teleportation vector $T$, as it does not appear
        in \eqref{eq:pn-simple-bh}.
\end{itemize}
Memory usage of PageRank is proportional to $2|E| + 3N + 1$, since the black
hole vector $B$ is not present, whilst the memory usage of Black Hole Metric is
proportional to $2|E| + 4N + 1$: they only differ by a factor of $N$.

\subsection{Proof of convergence} \label{ss:proof-conv}
In this section, we will prove that the underlying random walker process of the
Black Hole metric always converges.
First, let's consider the modified adjacency matrix $A'$. We know that it is
obtained from $A$ by adding a new node (the Black Hole) and by modifying the
arcs. It is a well-formed network nonetheless, and it is possible to evaluate
its PageRank. We can define the PageRank transition matrix $M^*$ for this
network as:
\[
    M^* = d (A' + S^*V^*) + (1 - d)T^*V^*
\]
where $S^*$ is the same as \eqref{eq:s-star} and it is the sink vector $S$ with
the addition of an extra sink, the entry of the Black Hole. The
teleportation vector $T^*$ is easilly constructed:
\begin{equation} \label{eq:t-star}
    T^* = \begin{pmatrix}
        T \\[0.3em]
        1
    \end{pmatrix}
\end{equation}
$V^*$ must be a non-negative $1 \times N + 1$ vector. The personalization
vector controls the per-node teleportation probability, but as long as $
\sum^{N+1}_{i=0} v^*_i = 1$, PageRank is guaranteed to converge no matter which
nodes get teleported to, so we can arbitrarilly choose $V^*$ as long as such
condition is met:
\begin{equation} \label{eq:v-star}
    V^* = \begin{pmatrix}
        V & 0
    \end{pmatrix}
\end{equation}
Given that the sum of the elements of $V$ is $1$, the sum of the elements of
$V^*$ is also $1$. Because of \eqref{eq:s-star} \eqref{eq:t-star}
\eqref{eq:v-star}, we rewrite $M^*$ as:
\[
    M^* = d \left[ A' + \begin{pmatrix}
        S \\[0.3em]
        1
    \end{pmatrix}
    \begin{pmatrix}
        V & 0
    \end{pmatrix} \right] + (1 - d) \begin{pmatrix}
        T \\[0.3em]
        1
    \end{pmatrix}
    \begin{pmatrix}
        V & 0
    \end{pmatrix} =
\]
\[
    = d \left[ \begin{pmatrix}
            \bar{A} & B \\[0.3em]
            \zeromat{1}{N} & 0
        \end{pmatrix} +
        \begin{pmatrix}
        SV & \zeromat{N}{1} \\[0.3em]
        V & 0
    \end{pmatrix} \right] + (1 - d) \begin{pmatrix}
        TV & \zeromat{N}{1} \\[0.3em]
        V & 0
    \end{pmatrix} =
\]
\[
    = \begin{pmatrix}
        d (\bar{A} + SV) & dB \\[0.3em]
        dV & 0
    \end{pmatrix} +
    \begin{pmatrix}
        (1 - d) TV &  \zeromat{N}{1} \\[0.3em]
        (1 - d) V & 0
    \end{pmatrix} = \begin{pmatrix}
        d(\bar{A} + SV) + (1 - d) TV & dB \\[0.3em]
        V & 0
    \end{pmatrix} =
\]
\[
	= \begin{pmatrix}
        \bar{M} & dB \\[0.3em]
        V & 0
    \end{pmatrix} = M'
\]
The last matrix is the definition of the transition matrix $M'$ for the
underlying random walker process of the Black Hole Metric of the network with
adjacency matrix $A$ and sink vector $S$.

In conclusion, if we choose $T^*$ and $V^*$ appropriately, the underlying
random walker process for the Black Hole Metric and PageRank is the same. The
conditions we set do not affect the generality of this statement, and since
PageRank is guaranteed to converge for every network, we can safely assume that
the Black Hole Metric converges as well regardless of the network structure.

\subsection{Rank equality theorem} \label{ss:rank-eq}
In this section, we present the theorem proving that Black Hole Metric is a
generalization of PageRank. Before discussing the theorem, however, we
introduce the lemma\eqref{lem:link-eq}.

\begin{link-equivalence} \label{lem:link-eq}
    If $B = \zeromat{N}{1}$ then $M = \bar{M}$.
\end{link-equivalence}

\begin{proof}
    If every entry in the $B$ vector is $0$ it follows that, $\forall i \in
    [1,N]$, we have from \eqref{eq:bh-weight}:
    \[
        \sum_{k=1}^{{out}_i } \frac{h_i - r_{ik}}{{out}_i (h_i - l_i)} = b_i
        = 0
    \]
    Given that $h_i \ge r_{ij}$ and $h_i > l_i$, since the denominator is
    always greater than $0$, the only way the summation can be $0$ is if $h_i =
    r_{ij} \enspace \forall k \in [1, {out}_i]$. Let's substitute $r_{ij}$ with
    $h_i$ in \eqref{eq:bh-arc-weight}:
    \[
        \bar{a}_{ij} = \frac{r_{ij} - l_i}{{out}_i (h_i - l_i)} =
        \frac{\cancel{h_i - l_i}}{{out}_i (\cancel{h_i - l_i})} =
        \frac{1}{{out}_i}
    \]
    and since $r_{ij} = h_i \enspace \forall j \in [1, {out}_i]$ we have that
    $a_{ij} = \frac{1}{{out}_i} = \bar{a}_{ij}$, so $A = \bar{A}$. According to
    the definitions of the two matrices $M$ and $\bar{M}$ we have that $\bar{M}
    - M = \bar{A} - A = 0$ so $\bar{M} = M$.
\end{proof}

It is interesting to note that if $B$ is all zeros, the arc weights are all the
same, which is obvious since we are assigning maximum score to each neighbour.
Knowing that the two matrices $M$ and $\bar{M}$ are the same when the black
hole effect is absent, we can easily prove that the values produced by applying
both PageRank and Black Hole Metric are the same.

\begin{rank-equality} [of rank equality]
    If every entry in the $B$ vector is $0$ then $p_b = 0$, and $P = \bar{P}$:
    \[
        B = \zeromat{N}{1}
        \quad
        \Rightarrow
        \quad
        \begin{cases}
            P = \bar{P} \\
            p_b = 0
        \end{cases}
    \]
\end{rank-equality}

\begin{proof}
Given that $B = \zeromat{N}{1}$ then, for the lemma \ref{lem:link-eq}, the
random walker \eqref{eq:rw-bh} becomes:
\[
	\begin{pmatrix}
		\bar{P}	\\[0.3em]
		p_b
	\end{pmatrix} =
	\begin{pmatrix}
		M^\transpose	& V^\transpose \\[0.3em]
		\zeromat{1}{N}	& 0
	\end{pmatrix}_\infty
	\begin{pmatrix}
		P_0	\\[0.3em]
	    0
	\end{pmatrix}
\]
Let's name $V_1 = \onemat{R}{1} \cdot V \in \mathbb{R}^{R \times N}$ and
calculate the $n$-th power of matrix $M'{}^T$:
\[
	\setlength\arraycolsep{2pt}
	\begin{pmatrix}
		M^\transpose	& V_1^\transpose \\[0.3em]
		\zeromat{1}{N}	& 0
	\end{pmatrix}^2 =
	\begin{pmatrix}
		M^\transpose	& V_1^\transpose \\[0.3em]
		\zeromat{1}{N}	& 0
	\end{pmatrix} \cdot
	\begin{pmatrix}
		M^\transpose	& V_1^\transpose \\[0.3em]
		\zeromat{1}{N}	& 0
	\end{pmatrix} =
	\begin{pmatrix}
		(M^\transpose)^2	& M^\transpose V_1^\transpose \\[0.3em]
		\zeromat{1}{N}		& 0
	\end{pmatrix}
\]
\[
	\setlength\arraycolsep{2pt}
	\begin{pmatrix}
		M^\transpose	& V_1^\transpose \\[0.3em]
		\zeromat{1}{N}	& 0
	\end{pmatrix}^3 =
	\begin{pmatrix}
		(M^\transpose)^2	& M^\transpose V_1^\transpose \\[0.3em]
		\zeromat{1}{N}		& 0
	\end{pmatrix} \cdot
	\begin{pmatrix}
		M^\transpose	& V_1^\transpose \\[0.3em]
		\zeromat{1}{N}	& 0
	\end{pmatrix} =
	\begin{pmatrix}
		(M^\transpose)^3	& (M^\transpose)^2 V_1^\transpose \\[0.3em]
		\zeromat{1}{N}	& 0
	\end{pmatrix}
\]
\[
	\hdots
\]
\[
	\setlength\arraycolsep{2pt}
	\begin{pmatrix}
		M^\transpose	& V_1^\transpose \\[0.3em]
		\zeromat{1}{N}	& 0
	\end{pmatrix}^n =
	\begin{pmatrix}
		(M^\transpose)^n	& (M^\transpose)^{n-1} V_1^\transpose \\[0.3em]
		\zeromat{1}{N}	& 0
	\end{pmatrix}
\]
the limit for $n \to \infty$ is:
\[
	\lim_{n \to \infty}
	\begin{bmatrix}
		(M^\transpose)^n	& (M^\transpose)^{n-1}V_1^\transpose \\[0.3em]
		\zeromat{1}{N}		& 0
	\end{bmatrix} =
	\begin{pmatrix}
		M^\transpose_\infty	& M^\transpose_\infty V_1^\transpose \\[0.3em]
		\zeromat{1}{N}		& 0
	\end{pmatrix}
\]
so we may write the random walker as:
\[
	\begin{pmatrix}
		\bar{P}	\\[0.3em]
		p_b
	\end{pmatrix} =
	\begin{pmatrix}
		M^\transpose_\infty	& M^\transpose_\infty V_1^\transpose \\[0.3em]
		\zeromat{1}{N}		& 0
	\end{pmatrix}
	\begin{pmatrix}
		P_0	\\[0.3em]
	    0
	\end{pmatrix} =
	\begin{pmatrix}
		M^\transpose_\infty P_0	\\[0.3em]
        0
	\end{pmatrix}
\]
hence:
\[
	\begin{cases}
		\bar{P} = M^\transpose_\infty P_0 = P \\
		p_b = 0
	\end{cases}
\]
because of \eqref{eq:rw-normal}.
\end{proof}

%% file: experiments.tex
\section[Experiments]{Experiments} \label{s:experiments} 
In this section, we present the results of the experiments using Black Hole
metric with synthetic networks and a real world network. The objective is to
study the behaviour of the Black Hole metric using different networks having
different size and different topology. While we expect the Black Hole Metric to
produce a different ranking, we make no claims that the produced ranking is an
improvement over the ranking produced by PageRank, as it is hard to generate or
find a network that allows us to clearly highlight the effect mentioned in the
toy example. Nonetheless, the possibility exists, and our metric still stands
as the only solution (to the best of our knowledge) to this hard to detect
issue.

In particular, we chose to present the results for six different synthetic
networks, three weighted directed Erd\H{o}s-–R\'enyi random graph networks
\cite{erdos:1959} of size 1000, 10000 and 100000, and three weighted directed
scale-free random networks, of size 1000, 10000 and 100000. The chosen networks
all differ either in size or in topology, and form a usable set of networks of
different characteristics. All Erd\H{o}s-–R\'enyi networks were created so that
the average outdegree is 10 and in addition all generated the directed
scale-free networks following the algorithm described by Bollob\'as in
\cite{bollobas:2003}. Using the same notation of~\cite{bollobas:2003},
we choose parameters for the generated networks as follows:
\medskip
\begin{center}
    \begin{tabular}{|c|c|l|}
        \hline
        Parameter & Value & Description \\
        \hline
        $\alpha$ & $0.41$ & Prob. of adding an edge from a new node to
        an existing one. \\
        $\beta$ & $0.54$ & Prob. of adding an edge between two existing
        nodes. \\
        $\gamma$ & $0.05$ & Prob. of adding an edge from an existing
        node to a new one. \\
        $\delta_{in}$ & $0.20$ & Bias for choosing nodes from in-degree
        distribution. \\
        $\delta_{out}$ & $0$ & Bias for choosing nodes from out-degree
        distribution. \\
        \hline
    \end{tabular}
\end{center}
\medskip
The idea behind the experiments we performed is to test the Black Hole metric
in a "wary" environment
to show
that PageRank does not make difference if the weights are all multiplied by a
constant factor. Therefore, assuming that the weight range for each arc is $[0,
99]$, the generated weights in each network were set to be in range $[0, 49]$,
the lower half of the full range. Then, we applied both PageRank and the Black
Hole Metric, and we derived the rank position of each node in the network. This
is the first step of the simulation. For the sake of convenience, we named the
two result sets respectively $PR_1$ and $BH_1$. After the first step, we
multiplied the weights by a factor of $\sfrac{99}{49}$, effectively scaling the
weights range from $[0, 49]$ to $[0, 99]$. We applied both metrics again and
named the result sets $PR_2$ and $BH_2$. This is the second step of the
simulation. As expected, we had $PR_1 = PR_2$, so, for ease of notation, we
will call the PageRank result set for both steps $PR$.

\begin{figure}[htbp]
    \centering
    \begin{minipage}[b]{0.36\textwidth}
        \includegraphics{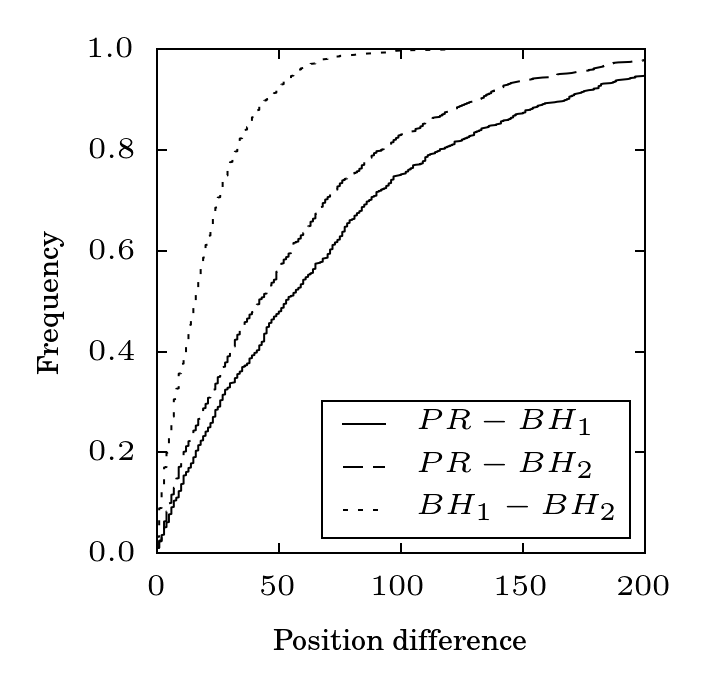}
        \subcaption{Network of size 1000}
    \end{minipage}%
    \begin{minipage}[b]{0.3\textwidth}
        \includegraphics{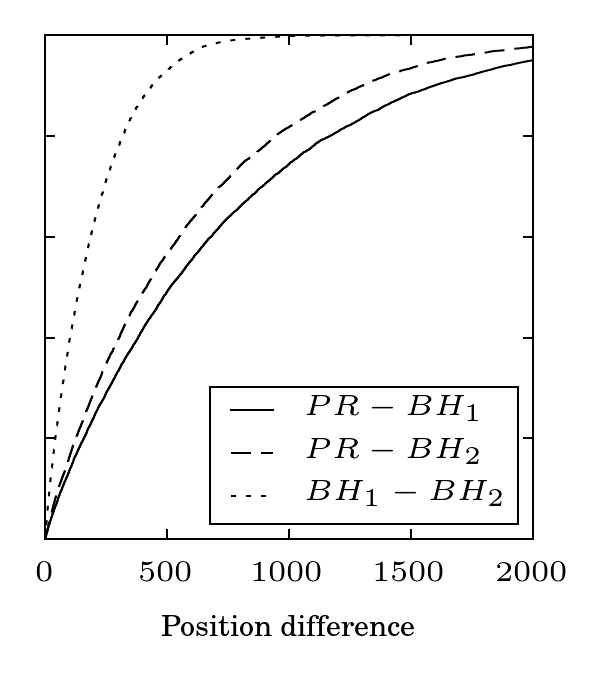}
        \centering
        \subcaption{Network of size 10000}
    \end{minipage}%
    \begin{minipage}[b]{0.3\textwidth}
        \includegraphics{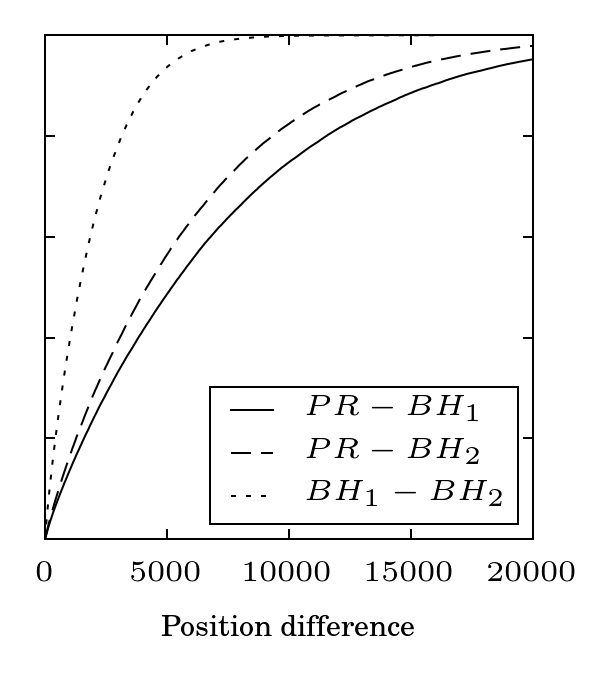}
        \subcaption{Network of size 100000}
    \end{minipage}
    \caption{Comparison of the empirical CDF of the absolute rank position
        difference (Erd\H{o}s-–R\'enyi networks)}
    \label{fig:er-cdf}
\end{figure}

\begin{figure}[htbp]
    \centering

    \begin{minipage}[b]{0.36\textwidth}
        \includegraphics{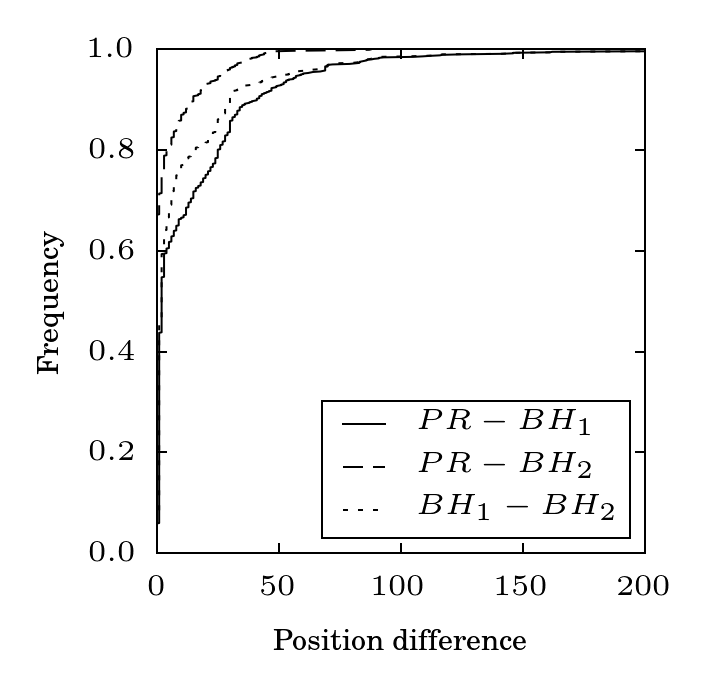}
        \subcaption{Network of size 1000}
    \end{minipage}%
    \begin{minipage}[b]{0.3\textwidth}
        \includegraphics{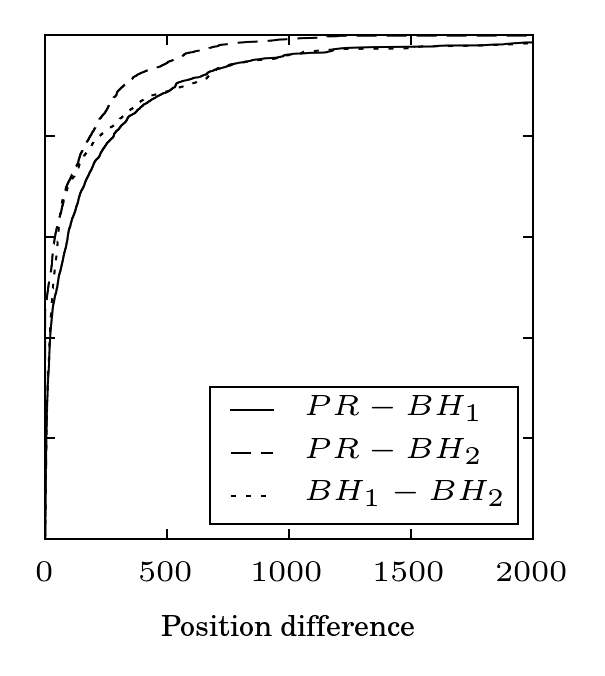}
        \subcaption{Network of size 10000}
    \end{minipage}%
    \begin{minipage}[b]{0.3\textwidth}
        \includegraphics{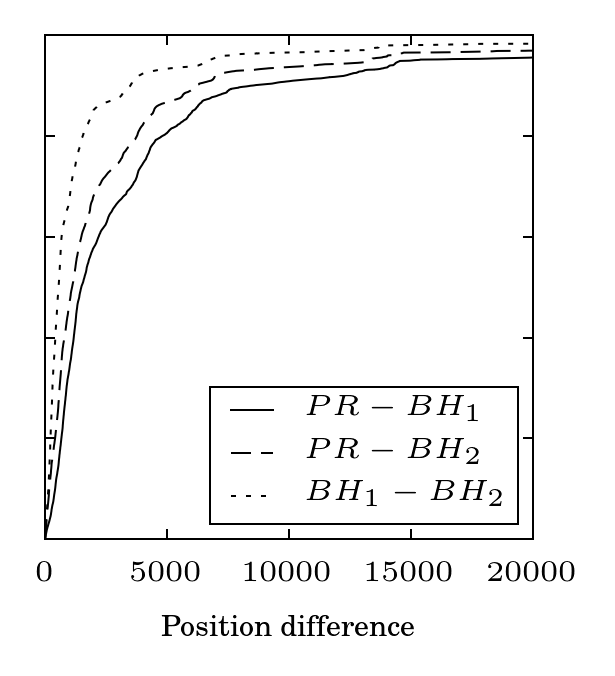}
        \subcaption{Network of size 100000}
    \end{minipage}

    \caption{Comparison of the empirical CDF of the absolute rank position
        difference (scale-free networks)}
    \label{fig:sf-cdf}
\end{figure}

\begin{figure}[htpb]
    \centering
    \begin{minipage}[b]{0.36\textwidth}
        \includegraphics{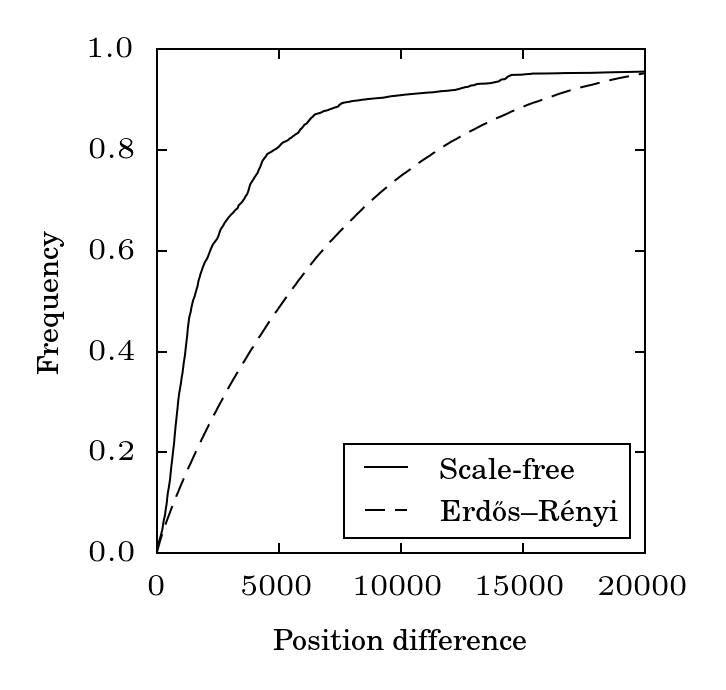}
        \subcaption{$PR - {BH}_1$}
    \end{minipage}%
    \begin{minipage}[b]{0.3\textwidth}
        \includegraphics{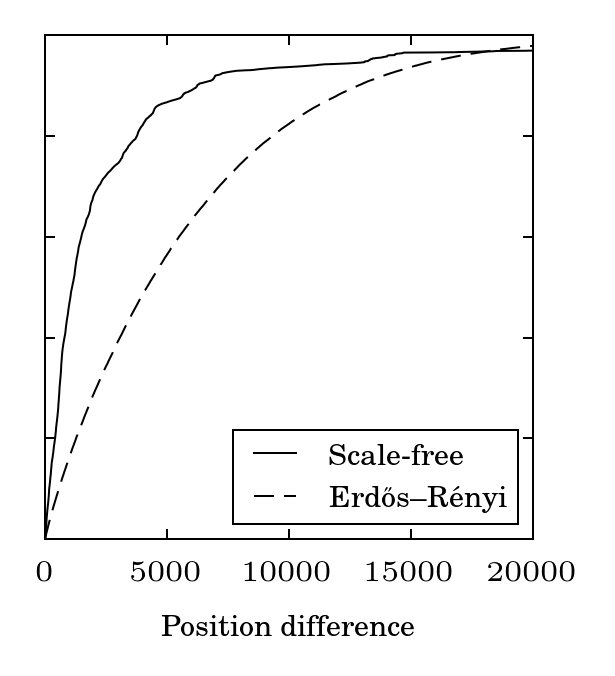}
        \subcaption{$PR - {BH}_2$}
    \end{minipage}%
    \begin{minipage}[b]{0.3\textwidth}
        \includegraphics{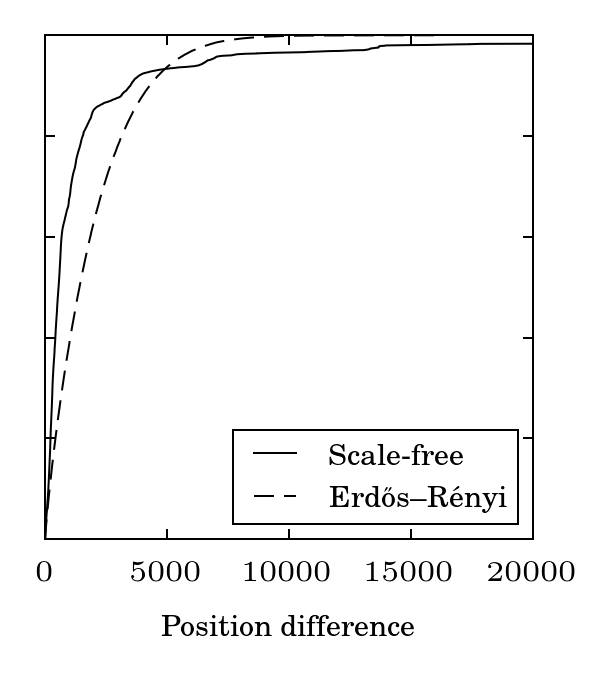}
        \subcaption{${BH}_1 - {BH}_2$}
    \end{minipage}
    \caption{Comparison of the empirical CDF of the absolute rank position
        difference, same size (100k nodes)}
    \label{fig:100k-cdf}
\end{figure}

The curves describe the cumulative distribution function of the absolute rank
position difference. We compared the result sets and condensed the results as
shown by Figures \ref{fig:er-cdf}, \ref{fig:sf-cdf} and \ref{fig:100k-cdf}. In
all the figures, the x-axis stands for the absolute rank position differences
between two results sets, while the y-axis stands for the cumulative frequency
of appearance. In order to better explain what the axes mean, let's take as an
example the solid line in \figref{fig:er-cdf}a, which depicts the frequency of
position difference between the result sets $PR$ and $BH_1$. We can see that
for a position difference of $50$ there is a frequency of about $0.4$. This
means that about $40\%$ of the nodes ranked in the result set $PR$ differ by at
most $50$ from the position they received in the result set $BH_1$. Since the
result sets are always compared pairwise, we will use the notation $R - Q$ to
illustrate the  absolute rank position difference among the result sets of $R$
and $Q$. To trim the outliers from the result sets, we have restricted the
x-axis to 20\% of the maximum possible rank position difference (which is equal
to the size of the network).

In \figref{fig:er-cdf} we show the results of both steps of the simulation of
the three Erd\H{o}s-–R\'enyi networks. Note that the network size does not
affect the shape of the curves; they are very similar for all three network
instances.  Moreover, the $PR - BH_2$ curve is always above the $PR - BH_1$
curve, which means that the $BH_2$ result set nearer to the $PR$ result set
than $BH_1$ is. This can be explained if we look at the arc weights: $BH_1$
comes from a network where the overall weights are lower than $BH_2$. In a
network with lower weights, the black hole steady-state probability is higher,
which means that it is more likely that a random walker, from any node, moves
to the black hole. But the black hole is a sink so the random walker will
teleport after reaching it. This means that the
black hole has an higher steady-state probability, and the teleportation effect
is amplified. 

As specified above, the PageRank result sets are identical in both simulation
steps meaning that the PageRank metric fails to capture the effect induced by
the different weight distribution. The dotted curve $BH_1 - BH_2$ highlights
that the two result sets $BH_1$ and $BH_2$ are always different. Note that this
curve is steeper than same curve related to the other two sets, because the
difference among the two result sets $BH_1$ and $BH_2$ is overall less than the
difference between either $BH_1$ or $BH_2$ and $PR$.

\figref{fig:sf-cdf} compares the result sets related to the three scale-free
network. The network size of scale-free networks does not influence much the
shape of the curves, however both $PR - BH_1$ and $PR - BH_2$ get smoother when
the network increases in size. Despite this difference, the curve $PR - BH_2$
keeps staying above the curve $PR - BH_1$, meaning that the Black Hole metric
assesses the difference in wariness of the nodes even in scale-free networks.
Finally, the two result sets $BH_1$ and $BH_2$ exhibit different behaviour when
the network size grows: the curve $BH_1 - BH_2$ is between the other two curves
when size is 1000, it almost coincides with $PR - BH_1$ when size is 10000, it
is above the other two curves when size is 100000.

At last, in \figref{fig:100k-cdf} we compare Erd\H{o}s-–R\'enyi and scale-free
networks of size 100000 by grouping together the curves of the same pair of
result sets. Note that the scale-free curves are different than the
Erd\H{o}s-–R\'enyi curves. This effect may depend on the different topology of
the two networks. In scale-free networks, nodes with high indegree, which are
few in number, are less affected by the weight fluctuations we introduced with
our experiments. Nodes with low indegree, which are more, are instead
strongly affected by the weight doubling, and their positions change a lot.
This causes the scale-free curves to appear steeper compared to the
Erd\H{o}s-–R\'enyi curves, although the behaviour of the Black Hole metric remains the same.



%

The second set of experiments we apply the Black Hole Metric to two real
world networks, Advogato and Libimseti.cz, retrieved from \cite{konect}.

Advogato (www.advogato.org) is an online community platform for free software
developers. As reported in the website of Advogato \emph{"Since 1999, our goal
has been to be a resource for free software developers around the world, and a
research testbed for group trust metrics and other social networking
technologies"}. Here we consider the Advogato \cite{konect:2016:advogato}
\cite{konect:massa09} trust network, where nodes are Advogato users and the
direct arcs represent trust relationships. Advogato names
\emph{"certification"} a trust link. There are three different levels of
certifications, corresponding to three weights for arcs: \emph{apprentice}
(0.6), \emph{journeyer} (0.8) and \emph{master} (1.0). A user with no trust
certificate is called an \emph{observer}. The network consists of 6541 nodes
and 51127 arcs and it exhibits an indegree and outdegree power law
distribution. As in the previously discussed experiments, we compute on this
network the PageRank and the Black Hole Metric and compare them using a
cumulative distribution graph, where the x-axis represents the possible
absolute rank position difference between the PageRank and the Black Hole
Metric of the nodes, while the y-axis represents the cumulative frequency of
appearance. To compute Black Hole Metric we set $l_i = 0.6$ and $h_i = 1.0$ for
all nodes in the network.

\begin{figure}[htpb]
    \centering
    \begin{minipage}[b]{0.5\textwidth}
        \includegraphics[width=\textwidth]{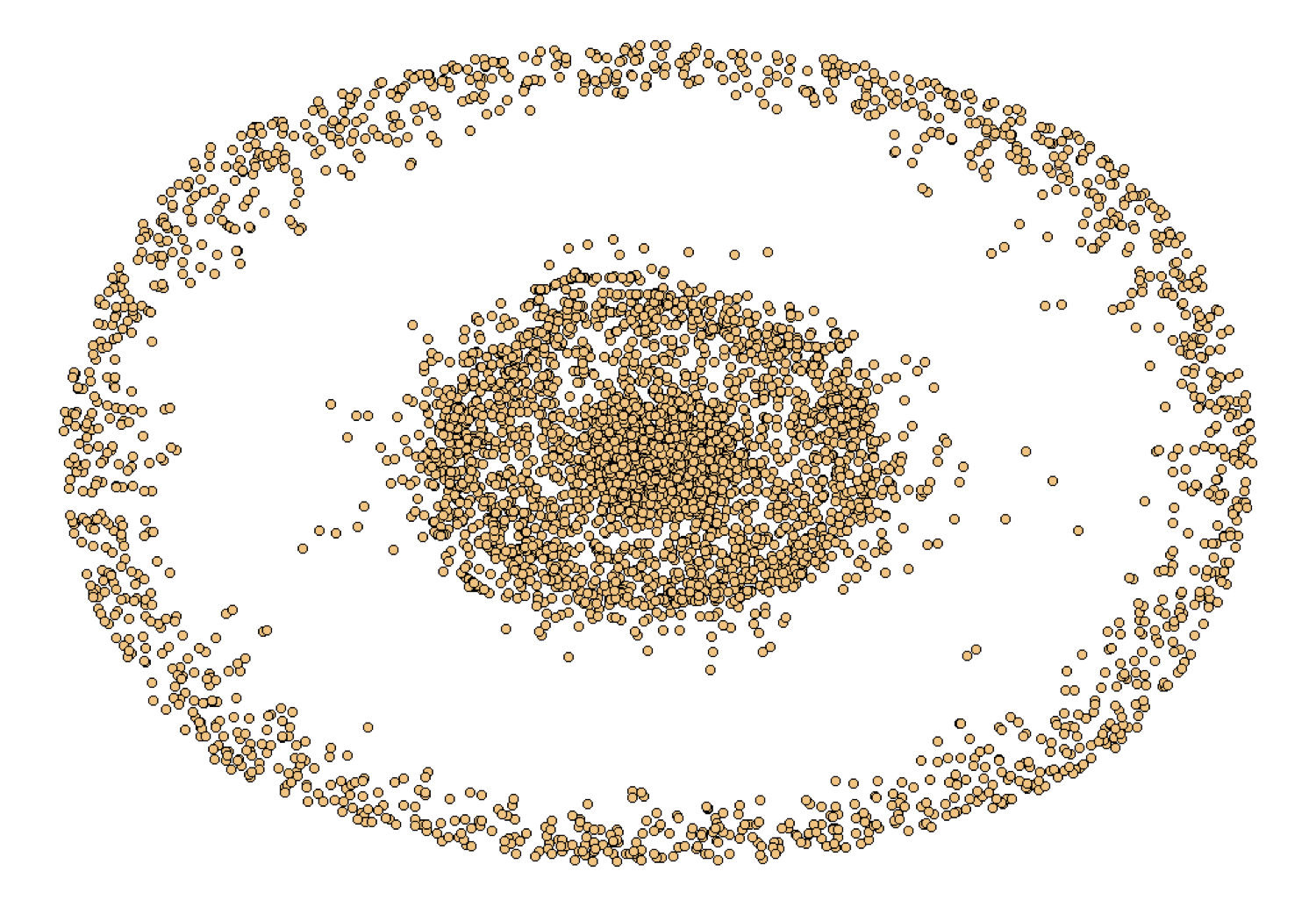}
        \subcaption{The network.}
    \end{minipage}%
    \begin{minipage}[b]{0.5\textwidth}
        \includegraphics[width=\textwidth]{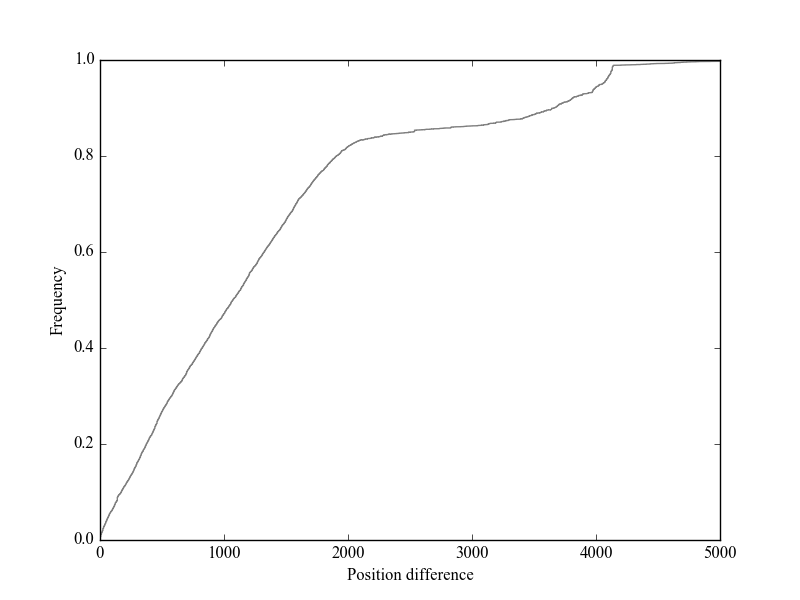}
        \subcaption{$PR - BH$ CDF.}
    \end{minipage}%
    \caption{The Advogato trust network}
    \label{fig:advogato-cdf}
\end{figure}

Figure~\ref{fig:advogato-cdf} shows the results. It is clear that the Black
Hole Metric produces different values (and ranks) compared to those computed by
using PageRank. In practice, it  means that Black Hole Metric produces a
different ranking compared to PageRank. For example, in
Table~\ref{tab:advogato}, we report the rank of the first 10 users of Advogato,
computed using the PageRank and Black Hole Metric.

\begin{table}[ht]
    \centering
    \begin{tabular}{|l|l|l|l|l|}
        \hline
        \multicolumn{1}{|c|}{\multirow{2}{*}{\textbf{Rank}}} &
        \multicolumn{2}{c|}{\textbf{PageRank}} &
        \multicolumn{2}{c|}{\textbf{BlackHole metric}} \\ \cline{2-5}
        \multicolumn{1}{|c|}{} &
        \multicolumn{1}{c|}{\textbf{NodeName}} &
        \multicolumn{1}{c|}{\textbf{PageRank Value}} &
        \multicolumn{1}{c|}{\textbf{NodeName}} &
        \multicolumn{1}{c|}{\textbf{BlackHole Value}} \\ \hline
        1  & federico   & 0.02093458    & alan      & 0.00594131 \\ \hline
        2  & alan       & 0.00978148    & miguel    & 0.00387012 \\ \hline
        3  & miguel     & 0.00658376    & rms       & 0.00290212 \\ \hline
        4  & raph       & 0.00405245    & raph      & 0.00230948 \\ \hline
        5  & rms        & 0.00381952    & federico  & 0.00176002 \\ \hline
        6  & jwz        & 0.00274046    & jwz       & 0.00172800 \\ \hline
        7  & davem      & 0.00262117    & rasmus    & 0.00158964 \\ \hline
        8  & rth        & 0.00258019    & rth       & 0.00158964 \\ \hline
        9  & rasmus     & 0.00250191    & gstein    & 0.00138078 \\ \hline
        10 & gstein     & 0.00230680    & davem     & 0.00135993 \\ \hline
    \end{tabular}
    \vskip10pt
    \caption{Rank of the first 10 users of Advogato trust network computed by
        using PageRank and Black Hole Metric.}
    \label{tab:advogato}
\end{table}

As reported in the website of Advogato, in order to assess the certification
level of each user they use a basic trust metric computed relatively to a
"seed" of trusted accounts. The original four trust metric seeds, set in 1999
when Advogato went online, were: $raph$ (Raph Levien), $miguel$ (Miguel Icaza),
$federico$ (Federico Mena-Quntero) and $alan$ (Alan Cox). In 2007 $mako$
(Benjamin Mako Hill) replaced $federico$. As we can infer from
Table~\ref{tab:advogato} both metrics are somewhat able to capture the
important role covered by the Advogato trust metric seeds, by putting them in
the top positions. However, Black Hole Metric, in our opinion, produces a more
appropriate ranking, according to the following observations:
\begin{itemize}
    \item $federico$ is first according to PageRank, while is 5th
        according to Black Hole Metric. Moreover the PageRank $federico$'s
        value is also significantly higher compared to $alan$ (the second in
        the chart), which means that $federico$ is steadly in the first
        position with a wide margin, despite the fact that he has not been a
        seed since 2007. We believe that lower position that the Black Hole
        Metric assigns to $federico$ better captures the fact that he was
        swapped out of the seed set.
    \item Another interesting difference is about the different position of the
        node $mako$. It is ranked 257$^{th}$ by the PageRank and 142$^{th}$ by
        Black Hole Metric. This ranking difference suggest that the Black Hole
        Metric better captures the relevance that $mako$ has been assuming
        inside the Advogato community.
\end{itemize}
%
%

%% file: conclusion.tex
\section[Conclusion]{Conclusion and future works} \label{s:conclusion}
In this article, we proposed a new PageRank based metric called Black Hole
Metric which aims at solving the normalization issue of PageRank algorithm. 
We provided examples that
highlight these problems and show that Black Hole Metric provides a different
ranking that takes into account the relative weights of the node outlinks. We
formally defined Black Hole Metric proving that it is an extension of PageRank.
We also compared the computational complexity of Black Hole Metric and PageRank
proving that they are quite similar. We proved that the Black Hole Metric
always converges. Finally, we experimented with the metric on several different
networks and showed first results, 
suggesting that the Black Hole metric seems to capture particular nodes behaviours;
this actually deserves further investigation in order to assess Black Hole Metric semantics,
and how it can help to model and address the ranking problem in different contexts.
In addition to this issue, others must be addressed. 

In paragraph \ref{ss:example} we said that the black hole rank value is
meaningless for the node ranking. Is the black hole just a mathematical trick
used to guarantee the stochasticity of the link matrix or does it have an
additional meaning? It is clear that the higher the PageRank of the black hole
is, the more the nodes of the network do not trust each other. It would be
interesting to study the possible correlation between the lack of trust of the
nodes and the position or value of the black hole in the Black Hole Metric
ranking.

Another open issue concerns the security of Black Hole Metric. We did not
investigate possible vulnerabilities of the metric, as they were not the focus
of this article. How does the Black Hole Metric behave in a network where
security considerations are important?  Does it guarantee protection against
common and uncommon attacks by internal or external agents
\cite{carchiolo:idc:2008}?

At last, in this article we did not investigate about methods and algorithms to
practically compute the steady-state probability vector. Black Hole Metric can
be seen as a generalization of PageRank and as such, many of the algorithms
that compute PageRank could be adapted to work with Black Hole Metric. It would
be interesting to find out to what extent is it possible to reuse existing
PageRank computation methods in order to improve the applicability of the Black
Hole Metric.

%% file: article.bbl
\begin{thebibliography}{10}

\bibitem{artz@2007}
Donovan Artz and Yolanda Gil.
\newblock A survey of trust in computer science and the semantic web.
\newblock {\em Web Semantics: Science, Services and Agents on the World Wide
  Web}, 5(2):58--71, 2007.

\bibitem{Bahmani:2011}
Bahman Bahmani, Kaushik Chakrabarti, and Dong Xin.
\newblock Fast personalized pagerank on mapreduce.
\newblock In {\em Proceedings of the 2011 ACM SIGMOD International Conference
  on Management of Data}, SIGMOD '11, pages 973--984, New York, NY, USA, 2011.
  ACM.

\bibitem{Bahmani:2010}
Bahman Bahmani, Abdur Chowdhury, and Ashish Goel.
\newblock Fast incremental and personalized pagerank.
\newblock {\em Proc. VLDB Endow.}, 4(3):173--184, December 2010.

\bibitem{bedi:2014}
Punam Bedi and Pooja Vashisth.
\newblock Empowering recommender systems using trust and argumentation.
\newblock {\em Information Sciences}, 279:569--586, 2014.

\bibitem{Berkhin:2005}
Pavel Berkhin.
\newblock A survey on pagerank computing.
\newblock {\em Internet Mathematics}, 2(1), 2005.

\bibitem{Bianchini:2005}
Monica Bianchini, Marco Gori, and Franco Scarselli.
\newblock Inside pagerank.
\newblock {\em ACM Trans. Internet Technol.}, 5(1):92--128, Febraury 2005.

\bibitem{bollobas:2003}
B{\'e}la Bollob\'{a}s, Christian Borgs, Jennifer Chayes, and Oliver Riordan.
\newblock Directed scale-free graphs.
\newblock In {\em Proceedings of the Fourteenth Annual ACM-SIAM Symposium on
  Discrete Algorithms}, SODA '03, pages 132--139, Philadelphia, PA, USA, 2003.
  Society for Industrial and Applied Mathematics.

\bibitem{pagerank:98}
S.~Brin and L.~Page.
\newblock The anatomy of a large-scale hypertextual web search engine.
\newblock In {\em Seventh International World-Wide Web Conference (WWW 1998)},
  1998.

\bibitem{Carchiolo20101893}
Vincenza Carchiolo, Alessandro Longheu, and Michele Malgeri.
\newblock Reliable peers and useful resources: Searching for the best
  personalised learning path in a trust- and recommendation-aware environment.
\newblock {\em Information Sciences}, 180(10):1893 -- 1907, 2010.
\newblock Special Issue on Intelligent Distributed Information Systems.

\bibitem{carchiolo:idc:2008}
Vincenza Carchiolo, Alessandro Longheu, Michele Malgeri, and Giuseppe Mangioni.
\newblock Trusting evaluation by social reputation.
\newblock In {\em Intelligent Distributed Computing, Systems and Applications,
  Proceedings of the 2nd International Symposium on Intelligent Distributed
  Computing - IDC 2008}, pages 75--84. Springer, 2008.

\bibitem{Carchiolo2012}
Vincenza Carchiolo, Alessandro Longheu, Michele Malgeri, and Giuseppe Mangioni.
\newblock {\em Gain the Best Reputation in Trust Networks}, pages 213--218.
\newblock Springer Berlin Heidelberg, Berlin, Heidelberg, 2012.

\bibitem{CPE:CPE1856}
Vincenza Carchiolo, Alessandro Longheu, Michele Malgeri, and Giuseppe Mangioni.
\newblock Trust assessment: a personalized, distributed, and secure approach.
\newblock {\em Concurrency and Computation: Practice and Experience},
  24(6):605--617, 2012.

\bibitem{Carchiolo:2013}
Vincenza Carchiolo, Alessandro Longheu, Michele Malgeri, and Giuseppe Mangioni.
\newblock Users attachment in trust networks: Reputation vs. effort.
\newblock {\em Int. J. Bio-Inspired Comput.}, 5(4):199--209, July 2013.

\bibitem{Carchiolo2015}
Vincenza Carchiolo, Alessandro Longheu, Michele Malgeri, and Giuseppe Mangioni.
\newblock Searching for experts in a context-aware recommendation network.
\newblock {\em Computers in Human Behavior}, 51:1086 -- 1091, 2015.
\newblock Computing for Human Learning, Behaviour and Collaboration in the
  Social and Mobile Networks Era.

\bibitem{Chen:2015}
Li~Chen, Guanliang Chen, and Feng Wang.
\newblock Recommender systems based on user reviews: The state of the art.
\newblock {\em User Modeling and User-Adapted Interaction}, 25(2):99--154, June
  2015.

\bibitem{Mcknight:1996}
N.~L.~Chervany D.~H.~McKnight.
\newblock The meanings of trust.
\newblock Technical report, Minneapolis, USA, 1996.

\bibitem{erdos:1959}
Paul Erd{\H o}s and Alfr{\'e}d R{\'e}nyi.
\newblock On random graphs i.
\newblock {\em Publicationes Mathematicae (Debrecen)}, 6:290--297, 1959 1959.

\bibitem{Gleich:2014}
David~F. Gleich.
\newblock Pagerank beyond the web.
\newblock {\em CoRR}, abs/1407.5107, 2014.

\bibitem{Gupta:2013}
Pankaj Gupta, Ashish Goel, Jimmy Lin, Aneesh Sharma, Dong Wang, and Reza Zadeh.
\newblock Wtf: The who to follow service at twitter.
\newblock In {\em Proceedings of the 22Nd International Conference on World
  Wide Web}, WWW '13, pages 505--514, Republic and Canton of Geneva,
  Switzerland, 2013. International World Wide Web Conferences Steering
  Committee.

\bibitem{kamvar:2003}
Sepandar~D. Kamvar, Mario~T. Schlosser, and Hector Garcia-Molina.
\newblock The {EigenTrust} algorithm for reputation management in {P2P}
  networks.
\newblock In {\em Proceedings of the Twelfth International World Wide Web
  Conference, 2003.}, 2003.

\bibitem{kim:2012}
Young~Ae Kim and Rasik Phalak.
\newblock A trust prediction framework in rating-based experience sharing
  social networks without a web of trust.
\newblock {\em Information Sciences}, 191:128--145, 2012.

\bibitem{kleinberg:1998}
Jon~M. Kleinberg.
\newblock Authoritative sources in a hyperlinked environment.
\newblock {\em Journal of ACM}, 46(5):604--632, 1999.

\bibitem{konect}
Jérôme Kunegis.
\newblock {KONECT} -- {The} {Koblenz} {Network} {Collection}.
\newblock In {\em Proc. Int. Conf. on World Wide Web Companion}, pages
  1343--1350, 2013.

\bibitem{Langville:2004}
Amy~N. Langville and Carl~D. Meyer.
\newblock Deeper inside pagerank.
\newblock {\em Internet Mathematics}, 1:2004, 2004.

\bibitem{lee:2003}
HyunChul Lee and Allan Borodin.
\newblock Perturbation of the hyper-linked environment.
\newblock In Tandy Warnow and Binhai Zhu, editors, {\em Computing and
  Combinatorics}, volume 2697 of {\em Lecture Notes in Computer Science}, pages
  272--283. Springer Berlin Heidelberg, 2003.

\bibitem{HyunChul:2003}
HyunChul Lee and Allan Borodin.
\newblock Perturbation of the hyper-linked environment.
\newblock In Tandy Warnow and Binhai Zhu, editors, {\em Computing and
  Combinatorics}, volume 2697 of {\em Lecture Notes in Computer Science}, pages
  272--283. Springer Berlin Heidelberg, 2003.

\bibitem{Lempel:2001}
R.~Lempel and S.~Moran.
\newblock Salsa: The stochastic approach for link-structure analysis.
\newblock {\em ACM Trans. Inf. Syst.}, 19(2):131--160, April 2001.

\bibitem{marsh:1994}
S.~Marsh.
\newblock {\em Formalising Trust as a Computational Concept}.
\newblock PhD thesis, 1994.

\bibitem{konect:massa09}
Paolo Massa, Martino Salvetti, and Danilo Tomasoni.
\newblock Bowling alone and trust decline in social network sites.
\newblock In {\em Proc. Int. Conf. Dependable, Autonomic and Secure Computing},
  pages 658--663, 2009.

\bibitem{Najork:2007}
Marc~A. Najork.
\newblock Comparing the effectiveness of hits and salsa.
\newblock In {\em Proceedings of the Sixteenth ACM Conference on Conference on
  Information and Knowledge Management}, CIKM '07, pages 157--164, New York,
  NY, USA, 2007. ACM.

\bibitem{konect:2016:advogato}
Advogato network~dataset {KONECT}, January 2016.

\bibitem{pagerank:99}
Lawrence Page, Sergey Brin, Rajeev Motwani, and Terry Winograd.
\newblock The pagerank citation ranking: Bringing order to the web.
\newblock Technical Report 1999-66, Stanford InfoLab, November 1999.
\newblock Previous number = SIDL-WP-1999-0120.

\bibitem{Prabha:2014}
S.~Prabha, K.~Duraiswamy, and J.~Indhumathi.
\newblock Comparative analysis of different page ranking algorithms.
\newblock {\em International Journal of Computer, Electrical, Automation,
  Control and Information Engineering}, 8(8):1486 -- 1494, 2014.

\bibitem{pretto:2002}
Luca Pretto.
\newblock A theoretical analysis of google's pagerank.
\newblock In AlbertoH.F. Laender and ArlindoL. Oliveira, editors, {\em String
  Processing and Information Retrieval}, volume 2476 of {\em Lecture Notes in
  Computer Science}, pages 131--144. Springer Berlin Heidelberg, 2002.

\bibitem{richardson:2002}
Mathew Richardson and Pedro Domingos.
\newblock The {I}ntelligent {S}urfer: {P}robabilistic {C}ombination of {L}ink
  and {C}ontent {I}nformation in {P}age{R}ank.
\newblock In {\em Advances in Neural Information Processing Systems 14}. MIT
  Press, 2002.

\bibitem{Roa-Valverde:2014}
Antonio~J. Roa-Valverde and Miguel-Angel Sicilia.
\newblock A survey of approaches for ranking on the web of data.
\newblock {\em Inf. Retr.}, 17(4):295--325, August 2014.

\bibitem{ruohomaa:2005}
Sini Ruohomaa and Lea Kutvonen.
\newblock Trust management survey.
\newblock In {\em proceedings of ITRUST 2005, number 3477 in LNCS}, pages
  77--92. Springer-Verlag, 2005.

\bibitem{Sargolzaei:2010}
P.~Sargolzaei and F.~Soleymani.
\newblock Pagerank problem, survey and future research directions.
\newblock 5(19):937--956, 2010.

\bibitem{Senanayake:2015}
Upul Senanayake, Mahendra Piraveenan, and Albert Zomaya.
\newblock The pagerank-index: Going beyond citation counts in quantifying
  scientific impact of researchers.
\newblock {\em PLoS ONE}, 10(8):e0134794, 08 2015.

\bibitem{serrano:2011}
Jesus Serrano-Guerrero, Enrique Herrera-Viedma, José Olivas, Andres Cerezo, and
  Francisco Romero.
\newblock A google wave-based fuzzy recommender system to disseminate
  information in university digital libraries 2.0.
\newblock {\em Information Sciences}, 181:1503--1516, 05 2011.

\bibitem{serrano:2013}
Jesus Serrano-Guerrero, Francisco Romero, and José Olivas.
\newblock Hiperion: A fuzzy approach for recommending educational activities
  based on the acquisition of competences.
\newblock {\em Information Sciences}, pages~--, 11 2013.

\bibitem{Ashutosh:2009}
Ashutosh~Kumar Singh and Ravi~Kumar P.
\newblock A comparative study of page ranking algorithms for information
  retrieval.
\newblock 3(4):745 -- 756, 2009.

\bibitem{Wang:2008}
Xuanhui Wang, Tao Tao, Jian-Tao Sun, Azadeh Shakery, and Chengxiang Zhai.
\newblock Dirichletrank: Solving the zero-one gap problem of pagerank.
\newblock {\em ACM Transactions on Information Systems}, 26(2):1--29, 2008.

\bibitem{Wenpu:2004}
Wenpu Xing and A.~Ghorbani.
\newblock Weighted pagerank algorithm.
\newblock In {\em Communication Networks and Services Research, 2004.
  Proceedings. Second Annual Conference on}, pages 305--314, May 2004.

\bibitem{Xiong:2004}
Li~Xiong and Ling Liu.
\newblock Peertrust: Supporting reputation-based trust for peer-to-peer
  electronic communities.
\newblock {\em IEEE Trans. on Knowl. and Data Eng.}, 16(7):843--857, July 2004.

\bibitem{zhirov:2010}
A.~O. Zhirov, O.~V. Zhirov, and D.~L. Shepelyansky.
\newblock Two-dimensional ranking of wikipedia articles.
\newblock {\em CoRR}, abs/1006.4270, 2010.

\bibitem{zhou:tpds:2007}
Runfang Zhou and Kai Hwang.
\newblock Powertrust: A robust and scalable reputation system for trusted
  peer-to-peer computing.
\newblock {\em IEEE Trans. Parallel Distrib. Syst.}, (4):460--473.

\bibitem{Zhu:2005}
Yangbo Zhu and Xing Li.
\newblock Distributed pagerank computation based on iterative
  aggregation-disaggregation methods.
\newblock In {\em Proceedings of the 14th ACM international conference on
  Information and knowledge management}, pages 578--585, 2005.

\end{thebibliography}
